\documentclass[twocolumn,aps, prb, showpacs,preprintnumbers,amsmath,amssymb]{revtex4}

\usepackage{graphicx}
\usepackage{dcolumn}
\usepackage{bm}
\usepackage[percent]{overpic}

\begin{document}
\newcommand{\Arg}[1]{\mbox{Arg}\left[#1\right]}
\newcommand{\bb}{\mathbf}
\newcommand{\braopket}[3]{\left \langle #1\right| \hat #2 \left|#3 \right \rangle}
\newcommand{\braket}[2]{\langle #1|#2\rangle}
\newcommand{\be}{\[}
\newcommand{\br}{\vspace{4mm}}
\newcommand{\bra}[1]{\langle #1|}
\newcommand{\braketbraket}[4]{\langle #1|#2\rangle\langle #3|#4\rangle}
\newcommand{\braop}[2]{\langle #1| \hat #2}
\newcommand{\dd}[1]{ \! \! \!  \mbox{d}#1\ }
\newcommand{\DD}[2]{\frac{\! \! \! \mbox d}{\mbox d #1}#2}
\renewcommand{\det}[1]{\mbox{det}\left(#1\right)}
\newcommand{\ee}{\]} 
\newcommand{\eg}{\textbf{\\  Example: \ \ \ }}
\newcommand{\Imag}[1]{\mbox{Im}\left(#1\right)}
\newcommand{\ket}[1]{|#1\rangle}
\newcommand{\ketbra}[2]{|#1\rangle \langle #2|}
\newcommand{\kp}{\arccos(\frac{\omega - \epsilon}{2t})}
\newcommand{\ldos}{\mbox{L.D.O.S.}}
\renewcommand{\log}[1]{\mbox{log}\left(#1\right)}
\newcommand{\Log}{\mbox{log}}
\newcommand{\Modsq}[1]{\left| #1\right|^2}
\newcommand{\nb}{\textbf{Note: \ \ \ }}
\newcommand{\op}[1]{\hat {#1}}
\newcommand{\opket}[2]{\hat #1 | #2 \rangle}
\newcommand{\occ}{\mbox{Occ. Num.}}
\newcommand{\Real}[1]{\mbox{Re}\left(#1\right)}
\newcommand{\so}{\Rightarrow}
\newcommand{\sol}{\textbf{Solution: \ \ \ }}
\newcommand{\thetafn}[1]{\  \! \theta \left(#1\right)}
\newcommand{\tin}{\int_{-\infty}^{+\infty}\! \! \!\!\!\!\!}
\newcommand{\Tr}[1]{\mbox{Tr}\left(#1\right)}
\newcommand{\kb}{k_B}
\newcommand{\rad}{\mbox{ rad}}
\preprint{APS/123-QED}

\title{Dynamic RKKY interaction in graphene}

\author{S. R. Power$^{(a)}$, F. S. M. Guimar\~aes$^{(b, c)}$, A. T. Costa$^{(d)}$, R. B. Muniz$^{(d)}$ and M. S. Ferreira$^{(a)}$}\email{ferreirm@tcd.ie}

\affiliation{%
(a) School of Physics, Trinity College Dublin, Dublin 2, Ireland \\
(b) Department of Physics and Astronomy; University of California, Irvine, California, 92697 U.S.A. \\
(c) CAPES Foundation, Ministry of Education of Brazil, Bras\'ilia/DF 70040-020, Brazil \\
(d) Instituto de Fisica Universidade Federal Fluminense, Brazil}%

\date{\today}

\begin{abstract}
The growing interest in carbon-based spintronics has stimulated a number of recent theoretical studies on the RKKY interaction in graphene, based on which the energetically favourable alignment between magnetic moments embedded in this material can be calculated. The general consensus is that the strength of the RKKY interaction in undoped graphene decays as $1/D^3$ or faster, where $D$ is the separation between magnetic moments. Such an unusually fast decay for a 2-dimensional system suggests that the RKKY interaction may be too short ranged to be experimentally observed in graphene. Here we show in a mathematically transparent form that a far more long ranged interaction arises when the magnetic moments are taken out of their equilibrium positions and set in motion. We not only show that this dynamic version of the RKKY interaction in graphene decays far more slowly but also propose how it can be observed with currently available experimental methods.

\end{abstract}

                 
\maketitle

\section{Introduction}
The atomically thin sheet of carbon known as graphene has been attracting the interest of the wider scientific community due to its enormous potential for applications in fields as diverse as photonics, sensor technology and spintronics, to name but a few \cite{riseofgraphene, neto:graphrmp, yazyev:review}. Spintronics is a particularly promising field for graphene application due to the weak spin-orbit and hyperfine interactions, which in other materials act as significant sources of spin relaxation and decoherence.\cite{kane_quantum_2005, huertas-hernando_spin-orbit_2006, min_intrinsic_2006, huertas-hernando_spin-orbit-mediated_2009, trauzettel_spin_2007, yazyev_hyperfine_2008, fischer_hyperfine_2009} 

One recurrent topic in the field of spintronics is the mechanism by which localized magnetic moments embedded into nanoscale systems are able to interact with each other even when far apart. An indirect exchange interaction mediated by the conduction electrons of a host medium manifests itself as an energy difference between different alignments of the localized moments, leading to energetically favourable configurations. Such an interaction is usually calculated within the Ruderman-Kittel-Kasuya-Yosida (RKKY) approximation\cite{RKKY:RK, RKKY:K, RKKY:Y} and indeed the interaction itself often takes this moniker. The RKKY interaction in graphene has been intensively studied \cite{Vozmediano:2005, dugaev:rkkygraphene, saremi:graphenerkky, brey:graphenerkky, hwang:rkkygraphene, bunder:rkkygraphene, black:graphenerkky, sherafati:graphenerkky, uchoa:rkkygraphene, black-schaffer_importance_2010, me:grapheneGF, sherafati:rkkygraphene2, kogan:rkkygraphene, disorderedRKKY} and there is a general consensus that its strength decays faster than in other 2-dimensional materials. In fact, the RKKY interaction between magnetic moments a distance $D$ apart decays as $1/D^3$ or faster in undoped graphene. Such a fast decay rate indicates that the interaction is rather short ranged and possibly not as easily observable as, for instance, in the case of carbon nanotubes \cite{vojislav:acs}, for which the RKKY interaction is predicted \cite{AntonioDavidIEC, David:IEC, DavidSpinValve} to decay as $1/D$. In the case of doped graphene\cite{obs} the RKKY interaction decays as $1/D^2$, as commonly expected for 2-dimensional materials, but it still remains to be experimentally seen.  

Rather than ruling out the RKKY in graphene as too short ranged an interaction to be seen in practice, here we show that its range is considerably augmented if the magnetic moments are set in motion. The notion of a dynamic RKKY has been proposed by  \v{S}im\'anek and Heinrich \cite{Simanek_gilbert_2003} who generalized the concept of the electron-mediated interaction between magnetic moments to the case of when these moments are not in static equilibrium. Moreover, an expanded range of the magnetic interaction has been reported in multilayered materials when their magnetization varies with time \cite{heinrich_dynamic_2003}, although no direct link with the RKKY interaction was established. In this manuscript we demonstrate that the magnetic interaction range in graphene can be expanded in a similar fashion. Mathematically transparent expressions for the dynamic spin susceptibility of this material are derived, based on which we identify a striking modification in the decay rate of the RKKY interaction when the magnetic moments are allowed to precess. 

\section{Model}
Let us start by recalling that the conventional RKKY interaction between two magnetic objects separated by a distance $D$ results from the energy involved in placing one of the objects in the presence of the electronic spin-polarization induced by the other. When minimizing this energy, the most favourable alignment of the magnetizations is found. It is thus not surprising that the strength of the RKKY decays with $D$ in exactly the same way as the induced spin polarization due to a single object. In other words, when determining the decay rate of the RKKY interaction, it is sufficient to consider how the induced spin polarization caused by a single magnetic object is spread in space. With that in mind, we shall first consider a single magnetic object on a graphene sheet. 
For the sake of simplicity, let us assume that the object is represented by a single substitutional magnetic atom. This assumption can be relaxed to consider more general magnetic objects without significantly affecting the conclusions obtained.

Instead of studying the effect of this magnetic atom in a static situation, we impose a perturbing time-dependent transverse magnetic field of the form
\begin{equation}
  \mathbf{h}_{\perp} = h_0 \; \left[ \, \cos(\omega \, t) \; \hat{\mathbf{x}} \; - \;  \sin(\omega \, t) \; \hat{\mathbf{y}}\, \right] 
\label{magfield}
\end{equation}
that sets the moment in precession with a frequency $\omega$. Experimentally this precession can be achieved in a number of different ways \cite{heinrich_dynamic_2003, tserkovnyak_nonlocal_2005, tombros_electronic_2007, costache_electrical_2008}. The spin polarization induced by the precessing moment becomes time dependent so that a second magnetic object would couple directly with the oscillatory polarization and precess in exactly the same way. This polarization is expressed by the transverse spin susceptibility $\chi$, which reflects how the spin degrees of freedom of a system respond to a magnetic excitation. 


To calculate the spin susceptibility one needs the Hamiltonian describing the electronic structure of the unperturbed system, which we assume is given by 
\begin{equation}
\hat{H} =   \sum_{<i,j>,\sigma} \gamma_{ij} \, \ {\hat c}_{i\sigma}^\dag \, {\hat c}_{j\sigma} + \sum_{\sigma} (\epsilon_0 \, {\hat n}_{0,\sigma} + {U \over 2}  \, {\hat n}_{0,\sigma} \, {\hat n}_{0,{\bar \sigma}}) + \hat{H}_Z \,.
\label{Hamiltonian}
\end{equation}
Here, $\gamma_{ij} $ represents the electron hopping between nearest neighbor sites $i$ and $j$,  $\hat{c}_{i\sigma}^{\dag}$ ($\hat{c}_{i\sigma}$) creates (annihilates) an electron with spin $\sigma$ in site $i$, $\epsilon_0$ is the atomic energy level of the magnetic site chosen to be located at the origin (site $0$), $\hat{n}_{0 \sigma} =
\hat{c}_{0,\sigma}^{\dag} \hat{c}_{0,\sigma}$ is the corresponding electronic
occupation number operator, and $U$ represents an effective on-site interaction
between electrons on the magnetic site, which is neglected elsewhere. Finally,
$H_Z$ plays the role of a local Zeeman interaction that defines the
$\hat{z}$-axis as the equilibrium direction of the magnetization. This unperturbed Hamiltonian is equivalent to that of the Anderson model used to describe localised magnetic impurity states in metals. \cite{anderson_localized_1961}. 
The interaction between the oscillatory magnetic field in Eq. \eqref{magfield} and the magnetization at site 0, $\mathbf{S}_0$, is accounted for by an interaction term $\hat{H}_{int} = g \mu_B \mathbf{h}_{\perp} \cdot \mathbf{S}_0$.
Our results are not critically dependent on the electronic structure parameters of the Hamiltonian, indicating that our findings are valid for a variety of magnetic objects. In fact, as far as the decay rate is concerned, it is fully determined by the host material, in this case the graphene sheet. Finally, spin-orbit coupling is neglected due to the long spin-diffusion length in graphene.

The time-dependent transverse spin susceptibility is defined as $\chi_{l,j}(t) = -{i \over \hbar} \Theta(t)\langle[{\hat S}_l^+(t),{\hat S}_j^-(0)]\rangle$, where $\Theta(x)$ is the Heaviside step function, and ${\hat S}_m^+$ and ${\hat S}_m^-$ are the spin raising and lowering operators at site $m$, respectively. The indices $j$ and $m$ refer to the locations where the field is applied and
where the response is measured, respectively. In our case, a precession of the magnetic moment is induced at the magnetic site $0$, and we wish to observe the spin disturbance a distance $D$ apart at an arbitrary site $m$. This response is fully described by $\chi_{m,0}(t)$. Within the random phase approximation, this susceptibility may be calculated in the frequency domain as 
\begin{equation}
\chi(\omega) = [1 + \chi^0(\omega)\, U]^{-1} \, \chi^0(\omega)
\label{RPAsus}
\end{equation}
where $\chi^0$ is the Hartree-Fock susceptibility, whose matrix elements are given by 
\begin{widetext}
\begin{equation}
\chi_{\ell,j}^0(\omega) = {i \hbar \over 2 \pi} \int_{-\infty}^{+\infty} dE^\prime f(E^\prime) \left\{ \left[G_{j,\ell}^\uparrow(E^\prime) -  G_{j,\ell}^{-\uparrow}(E^\prime)  \right]  G_{\ell,j}^\downarrow(E^\prime+\hbar \omega) +  \left[G_{\ell,j}^\downarrow(E^\prime) -  G_{\ell,j}^{-\downarrow}(E^\prime)  \right]  G_{j,\ell}^{-\uparrow}(E^\prime-\hbar \omega) \right\}\,\,.
\label{susc-hf}
\end{equation}
\end{widetext}
Here, $G_{\ell,j}^{\sigma}(E^\prime)$ and $G_{\ell,j}^{-\sigma}(E^\prime)$ represent the time Fourier transforms of the retarded and advanced single-electron Green Functions (GF), respectively, for an electron of energy $E^\prime$ with spin $\sigma$ between
sites $\ell$ and $j$, and $f(E^\prime)$ is the Fermi function. Concerning the physical meaning of the quantity $\chi_{m,0}(\omega)$, its absolute value is proportional to the amplitude of the spin precession at site $m$ as a response to a precession induced at site $0$. Therefore, to assess how a second magnetic object inserted at site $m$ will couple with that located at the origin, the relevant matrix element of the susceptibility is $\chi_{m,0}$. 
To simplify the analytic calculation of the interaction behaviour we consider the case of a single impurity. However, the approach can be generalised to the cases of magnetic clusters or multiple impurities by introducing a sum over impurity sites in the second term of the Hamiltonian given by Eq. \eqref{Hamiltonian}. Such an approach is used in numerical calculations involving more than one impurity later in this work. 

From Eqs. \eqref{RPAsus} and \eqref{susc-hf} it is clear that the position dependence of the spin susceptibility is entirely contained within the corresponding Hartree-Fock term $\chi^0_{m,0}$. Furthermore, the dependence of this term on the separation $D$ between the sites $0$ and $m$ can be calculated more easily by replacing the matrix elements of $G^{\sigma}(E^\prime)$ that appear in Eq.(\ref{susc-hf}) with the bulk GF of pristine graphene, which we calculate within the tight-binding formalism. In this case $\chi_{m,0}^0(\omega)$ is written as $\chi_{m,0}^{0} (\omega) = I_1 + I_2 + I_3$, where \cite{muniz_theory_2002}
\begin{equation}
\begin{split}
I_1 & = \frac{i}{2\pi} \int_{-\infty}^{+\infty}  dE^\prime f(E^\prime) {\cal G}_{0,m}(E^\prime){\cal G}_{m,0}(E^\prime+\hbar \omega)\\ 
I_2 & = - \frac{i}{2\pi} \int_{-\infty}^{+\infty}  dE^\prime f(E^\prime) {\cal G}^{-}_{m,0}(E^\prime){\cal G}^{-}_{0,m}(E^\prime-\hbar \omega)\\ 
I_3 & = \frac{i}{2\pi} \int_{-\infty}^{+\infty}  dE^\prime \left[ f(E^\prime + \hbar \omega) - f(E^\prime) \right] {\cal G}^{-}_{0,m}(E^\prime){\cal G}_{m,0}(E^\prime+\hbar \omega) \,\,.
\end{split}
\label{chi-approx}
\end{equation}
Here ${\cal G}$ (${\cal G}^-$) is the retarded (advanced) GF of pristine graphene. Note that $I_1$ ($I_2$) involves the convolution of two retarded (advanced) GF whereas $I_3$ involves one of each type.

We can now make use of the simplicity of the electronic structure of graphene and write the off-diagonal matrix elements of the pristine retarded GF of graphene as 
\begin{equation}
{\cal G}_{0,m}(E) = {\cal G}_{m,0}(E)={{\cal A}(E) \, e^{i {\cal Q}(E) D} \over \sqrt{D}}\,\,.
\label{g-pristine}
\end{equation}
This expression for the off-diagonal GF is valid for moderately large values of $D$. \cite{me:grapheneGF} Furthermore, it is valid for energies not only around the Fermi level where the dispersion relation is linear but also across the entire energy band. The precise forms of the functions ${\cal A}(E)$ and ${\cal Q}(E)$ depend on the direction of the vector joining the sites $0$ and $m$ on the graphene sheet but what is obvious from the expression is the simple functional form of ${\cal G}_{0,m}$ as a function of the separation $D$. For the case where the vector joining the sites $0$ and $m$ is along the armchair direction and both sites are located on the same sublattice, we can write
\begin{equation}
{\cal Q}(z) = \pm \sin^{-1}\left(\frac{z}{\gamma}\right)
\end{equation} and
\begin{equation}{\cal A}(z) = -i \sqrt{\frac{2}{i\pi}} \sqrt{\frac{z}{\left(z^2+3\gamma^2\right)\sqrt{\gamma^2-z^2} } } \,,
 \end{equation}
where $\gamma$ is the nearest neighbour hopping of graphene. In the next section we investigate how the amplitude of the dynamic susceptibility varies as the separation is increased along this direction. Changing the direction investigated may introduce additional oscillatory features, similar to those previously noted in the static case, but will not affect the decay rate of the interaction, which is our primary concern in this work. Allowing impurities to locate on different sublattices leads to qualitative changes in the static interaction where the sign of the coupling indicates preferred magnetic alignments. However, once again, the decay rate is not altered. This will be seen in Section \ref{expsig}, where multiple impurities located on both sublattices and separated in random directions are considered numerically to reproduce a more likely experimental configuration.

\section{Interaction range}

We can now investigate the decay rate of the RKKY interaction by studying the three integrals in Eq. \eqref{chi-approx} separately. By symmetry, $I_1$ and $I_2$ have similar functional forms and therefore identical contributions. For the sake of conciseness, let us focus on the integral $I_1$. Combining Eqs. (\ref{chi-approx}) and (\ref{g-pristine}) we may write
\begin{equation}
I_1  \sim \int_{-\infty}^{+\infty}  \mathrm{d} E^\prime \, \, \frac{{\cal B} (E^\prime,\omega) \, e^{ i( {\cal Q}(E^\prime) + {\cal Q}(E^\prime+\hbar \omega)) D}}{D (1 + e^{\beta(E^\prime - E_F)})} \,\,,
\label{J_integral}
\end{equation}
where $E_F$ is the Fermi energy, ${\cal B} (E^\prime,\omega) = {\cal A}(E^\prime) \times {\cal A}(E^\prime+\hbar \omega)$, $\beta = \frac{1}{k_B T}$, T being the temperature and $k_B$ the Boltzmann constant. The integral in Eq (\ref{J_integral}) can be solved by employing a semi-circular contour in the upper-half of the complex energy plane to reduce the integral to a sum over Matsubara frequencies. Expanding the functions ${\cal B}(E^\prime,\omega)$ and ${\cal Q}(E)$ around $E_F$ and taking the low temperature limit, $T \rightarrow 0$,  we find that $I_1$ is given by \cite{me:grapheneGF}

\begin{equation}
\label{I1final}
\begin{split}
I_1 =&\frac{1}{2\pi}e^{i[2{\cal Q}(E_F)+{\cal Q}^\prime(E_F) \hbar \omega] D}\sum_\ell \frac{(-1)^{\ell+1}{\cal B}^{\ell}(E_F, 0 )}{[2i{\cal Q}^\prime(E_F)]^{\ell+1}}\frac{1}{D^{\ell+2}}\ ,
\end{split}
\end{equation}
where $\ell$ is an integer and ${\cal B}^{\ell}(E_F,0)$  is the corresponding $\ell^{th}$-order derivative of the function ${\cal B}(E^\prime,0)$ evaluated at $E_F$, which results from its Taylor expansion. 

The $D$-dependence of $I_1$ is now evident. In the asymptotic limit\cite{David:IEC, me:grapheneGF} of large separations it is determined by the leading term in Eq. \eqref{I1final}, namely $\ell=0$, suggesting that, in general, $I_1 \sim D^{-2}$. However, at $E_F =0$, the coefficient ${\cal B}^{(0)}$ vanishes and the decay rate is in fact determined by the first surviving term, $\ell=1$, resulting in $I_1 \sim D^{-3}$ for undoped graphene. When $E_F \ne 0$, ${\cal B}^{(0)}$ does not vanish, and in this case $I_1$ does decay as $D^{-2}$. Notice that the frequency-dependence is contained only in the argument of the exponential in Eq.(\ref{I1final}), meaning that even when $\omega=0$ the decay rate of $I_1$ remains unaltered. This is precisely what is found for the static RKKY and agrees with the aforementioned feature that this interaction decays as $D^{-3}$ ($D^{-2}$) for undoped (doped) graphene. 

Regarding the $I_3$ integral, for $k_BT\rightarrow0$, the step-like Fermi functions can be used to rewrite this term as 
\begin{equation}
 I_3=- \frac{i}{2\pi} \int_{E_F-\hbar \omega}^{E_F} dE^\prime {\cal G}^*_{0,m}(E^\prime){\cal G}_{0,m}(E^\prime+\hbar \omega) \,.
\end{equation}
The pristine Green function ${\cal G}_{0,m}$ given by Eq.(\ref{g-pristine}) can be more concisely expressed for the small integration range around $E_F$  if one rewrites $
{\cal Q}(z) \simeq \frac{z}{\gamma}$ and ${\cal A}(z) \simeq -i\sqrt{\frac{2}{i\pi}}\frac{\sqrt{z}}{\sqrt{3|\gamma|^3}}$, respectively.
In this case, the distance dependence can be taken outside the integral and we find
\begin{equation}
 I_3=-\frac{1}{3 \pi D |\gamma|^3} e^{\pm i\frac{\hbar \omega}{\gamma}D} \int_{E_F-\hbar \omega}^{E_F} \, dE^\prime \, \sqrt{-E^\prime (E^\prime + \hbar \omega)} \,,
\label{i3+i4}
\end{equation}
which in the undoped ($E_F=0$) case reduces to 
\begin{equation}
I_3=-\frac{\hbar^2 \omega^2}{24\pi D |\gamma|^3} e^{\pm i\frac{\hbar \omega}{\gamma}D}\ .
\end{equation}
The separation dependence coming from $I_3$ is clearly more long ranged than that from $I_1$ and $I_2$ since Eq.(\ref{i3+i4}) decays as $D^{-1}$. Furthermore, unlike the expression for $I_1$ in Eq.(\ref{I1final}) which remains finite even in the static limit, the contribution from $I_3$ vanishes when $\omega=0$. Therefore, the RKKY interaction between localized magnetic moments in graphene becomes more long ranged once the moments are set in motion with a finite excitation frequency. 


\begin{figure}
\includegraphics[width=0.5\textwidth]{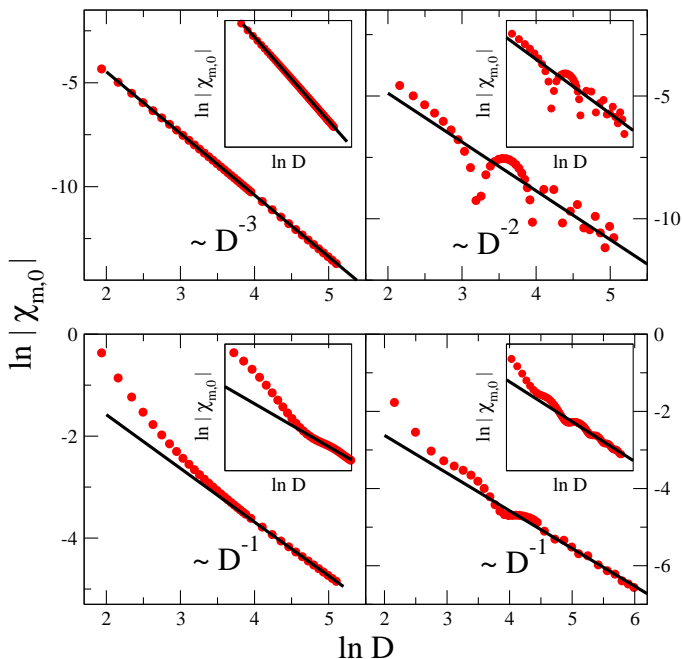}
\caption{Log-log plots of the numerically calculated spin susceptibility (in arbitrary units) probed at a distance $D$ from a single magnetic impurity. Top panels correspond to the case of static susceptibility whereas bottom panels are for finite frequencies. Left (Right) panels are for undoped (doped) cases. (Red) Dots are the numerically calculated values and (black) lines display the predicted power laws also highlighted on each of the main panels. Insets display the corresponding calculations for two impurities where exactly the same power law behavior is found.}
\label{figure1}
\end{figure}

Figure 1 confirms the analytically predicted decay rates with a fully numerical evaluation of the spin susceptibility using Eqs. \eqref{RPAsus} and \eqref{susc-hf}. The top panels depict $\vert \chi_{m,0} \vert$ for a single magnetic impurity located at site $0$ in a purely static configuration, {\it i.e.} $\omega=0$, whereas the bottom panels show the same result for the case in which the moment is precessing at the resonance frequency. Left (right) panels correspond to undoped (doped) cases. (Red) Dots are the numerically evaluated results and the (black) lines follow the power laws shown in the Figure. In the static case the spin polarization induced by the magnetic object decays as expected, at the same rate as the reported RKKY interaction that goes as $D^{-3}$ ($D^{-2}$) when $E_F=0$ ($E_F\neq0$). Attention is drawn to the $1/D$ asymptotic decay that appears in all dynamic cases and which indicates that the RKKY should indeed be more long ranged when the magnetic moments are in motion. This behavior is further confirmed by the respective insets which correspond to calculations of the spin susceptibility for the case of two magnetic impurities a distance $D$ apart. For these calculations the susceptibility in Eq. \eqref{susc-hf} uses numerical Green functions whose corresponding Hamiltonian includes the additonal terms necessary to describe a second magnetic impurity. In this case the spin disturbance at the second moment induced by the precession of the first is calculated. Note that the respective decay rates in this case are in excellent agreement with the predictions made for a single impurity, confirming the analytic treatment of the decay rate behaviour outlined above.

\section{Experimental signatures}
\label{expsig}
To date it has been very difficult to probe the RKKY interaction in graphene experimentally. It is understandable that with a decay rate as fast as $D^{-3}$ it is difficult to probe the interaction for any reasonable separation. The presence of magnetism in disordered graphene systems may indicate the presence of an exchange coupling between magnetic moments formed around defects. Nuclear magnetic resonance experiments reveal that these defects have indeed magnetic moments, since they couple to implanted Fe atoms \cite{sielemann_magnetism_2008}. However, whether or not these moments couple with each other, or with the graphene lattice, to form a ferromagnetic state is a controversial subject and many of the results in this area have proved difficult to reproduce \cite{yazyev:review}. 
However, our results suggest that magnetic moments should be able to feel their mutual presence at greater separations once they are set in motion. To address the issue of how this dynamic RKKY interaction can be probed, we turn our attention to a slightly different feature that can also be extracted from the spin susceptibility, namely the lifetime of magnetic excitations, something that
can be probed using the method of inelastic scanning tunneling spectroscopy (ISTS). \cite{heinrich_single-atom_2004, loth_controlling_2010, khajetoorians_detecting_2010,  khajetoorians_itinerant_2011}
Rather than looking at the off-diagonal element of the susceptibility for a given frequency, the diagonal matrix element $\chi_{m,m}$ plotted as a function of the frequency $\omega$ tells us how strongly the system responds to a time-dependent magnetic excitation. In particular, peaks in the $\omega$-dependent susceptibility reflect the existence of resonance frequencies whereas their inverse widths characterize the respective lifetime of the spin excitations. Figure 2A shows $\chi_{0,0}$ as a function of the excitation frequency $\omega$ for a single magnetic impurity at the origin (site $0$), where a single peak is clearly identifiable with a linewidth $W_1$. 

\begin{figure}
\includegraphics[width=0.45\textwidth]{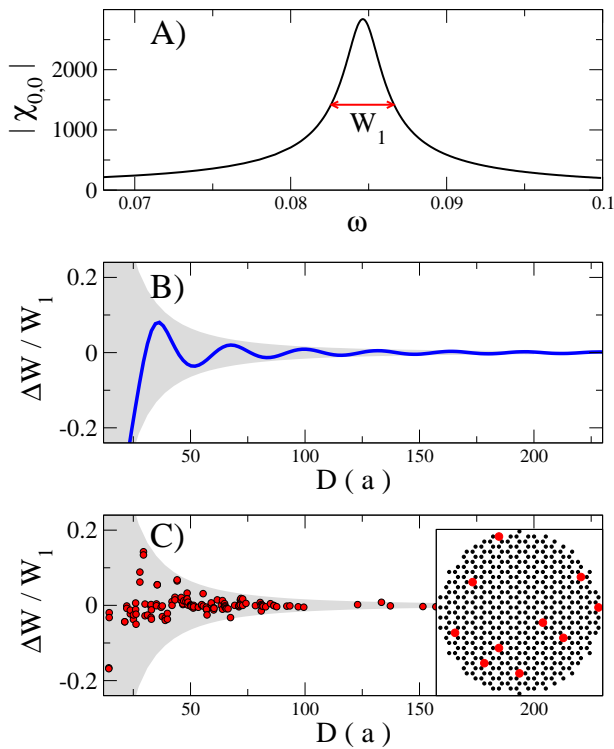}
\caption{(A) Diagonal element of the spin susceptibility $\chi_{0,0}$ (in arbitrary units) obtained for a single magnetic impurity located at site $0$  plotted as a function of the excitation frequency $\omega$. A single peak of width $W_1$ is highlighted. (B) Fractional deviations of the linewidth ($(W-W_1)/W_1$) plotted as a function of the relevant separation (in units of the lattice parameter of graphene). The solid (blue) line corresponds to the case of two impurities a distance $D$ apart. The gray area delimits the region spanned by the two-impurity result if all possible directions are included. (C) The scattered (red) dots correspond to the linewidths $W$ of the susceptibility $\chi_{m,m}$ for different sites $m$ in the case of a disordered array of magnetic impurities randomly distributed on graphene, as shown schematically in the inset, plotted as a function of the corresponding nearest neighbor separation, $D_{NN}$, i.e. the distance to the impurity nearest $m$ in the configuration.}
\label{figure2}
\end{figure}

To examine how the dynamic RKKY interaction between moments affects the lifetime of the spin excitations, a second moment is introduced into the system as before and its distance, $D$, from the first moment varied along the armchair direction. We have seen clearly in the previous section how the interaction between precessing moments manifests itself in the off-diagonal term of the spin susceptibility. It is now worth investigating how the diagonal term is affected since the resultant variations in the spin excitation spectrum may provide experimentally detectable signatures of a dynamic RKKY interaction. We expect that when the moments are moved far enough apart they become essentially independent and the excitation linewidth should approach that of the single impurity case. In fact, we can use Eq. \eqref{RPAsus} to expand the diagonal term of the susceptibility for two magnetic impurities at sites $0$ and $m$, $\chi^{(2)}_{0,0}$, as
\begin{equation}
 \chi^{(2)}_{0,0}  \approx \chi^{(1)}_{0,0} \, - \, \frac{\chi^0_{0,m} U \chi^0_{m,0} }{\left(1 + \chi^0_{0,0} U\right) \, \left(1 + \chi^0_{m,m} U\right)} \, - \, \cdots \,,
\end{equation}
where $\chi^{(1)}_{0,0}$ is the single-impurity susceptibility. The first-order correction to the diagonal matrix element of the single-impurity susceptibility when a second moment is introduced at site $m$ contains the product $\chi^0_{0,m}  \chi^0_{m,0}$ and should therefore decay as $D^{-2}$, with higher order terms in the expansion decaying even faster. This analysis suggests that fluctuations in the magnetic excitation spectrum induced by the introduction of an additional moment should decay as $D^{-2}$ towards the single-impurity result. 

Our numerical simulations find that adding a second impurity to the system maintains the peaked features in the spin susceptibility found for the single impurity case in  Figure 2A, but that now the linewidths become $D$-dependent. This can be viewed in Figure 2B where $\Delta W / W_1$ represents the relative deviation of the linewidth $W$ for the two impurity case from that of the single-impurity linewidth, $W_1$. The solid (blue) line shows that the linewidth oscillates around the value for the one-impurity case, with an amplitude that decays asymptotically towards it as the separation between the two moments is increased. This decay rate is found to be $D^{-2}$, as predicted above and indicated by the shaded area enveloping the two-impurity curve. It is worth noting that a mere phase shift in the solid line occurs if one changes the direction along which the two impurities are separated. If all possible directions were considered, the resulting data points would generate the shaded area in Figures 2B and 2C. 

Rather than having one or two isolated impurities,  a more realistic scenario is the case depicted in the inset of Figure 2C which shows several magnetic atoms randomly located across the graphene sheet. Many disordered configurations of impurities are generated where the impurities are allowed to locate in arbitrary directions from each other and occupy either sublattice. For each magnetic impurity we calculate the linewidth of the diagonal spin susceptibility and also record the separation from the nearest other impurity in that configuration. These linewidths are plotted as relative deviations as before, this time as a function of nearest-neighbour distance, $D_{NN}$, and appear in the main panel of Figure 2C as the solid (red) scattered dots. Notice that the spread of calculated linewidths for the disordered configurations agrees well with the shaded area that enveloped the two-impurity linewidth oscillation. Therefore, the dynamic interaction between the nearest-neighbour impurities tends to dominate and determine the overall lifetime of the spin excitation in these disordered configurations. Consequently, the fluctuations in the linewidth measurements reflect the variations of the dynamic RKKY interaction and the spread of measurements for a given distance, i.e. the shaded area, decays as $D^{-2}$. Bearing in mind that the nearest-neighbour separation is directly related to the concentration of impurities $\rho \sim D_{NN}^{-2}$, we predict that the standard deviation ($\sigma_W$) of the measurements of the spin-excitation lifetimes will scale with the concentration of magnetic impurities as $\sigma_W \sim \rho$. This is an unmistakable signature of the RKKY interaction in its dynamic form. Note that the inset of Figure 2C matches the experimental setup of Reference \cite{khajetoorians_itinerant_2011} which measured the spin excitation lifetimes of randomly dispersed Fe atoms on Cu surfaces with ISTS. It should be equally possible to probe the dynamic RKKY interaction in graphene using similar techniques since C atoms have substantially smaller spin-orbit coupling than Cu. We note that our studies have assumed low concentrations of magnetic impurities and also atomically precise graphene samples, which may be difficult to achieve experimentally. A recent study suggests that, for the static RKKY, strong on-site disorder can induce an exponential suppression of the coupling at large distances \cite{disorderedRKKY}. We expect that similar features may occur in the dynamic RKKY under the same kind of disorder regimes.

\section{Conclusion}

In summary, we have shown that the RKKY interaction between localized magnetic moments embedded in a non-magnetic material, which is predicted to decay rather fast in graphene, may become substantially more long ranged once the magnetic moments are taken out of equilibrium and set to precess. This has been illustrated using analytic arguments and fully numerical calculations. We argue that the difficulty of experimentally probing the RKKY interaction in graphene-based materials may be overcome by simply exciting the magnetization of magnetic objects in contact to graphene, something that is currently achievable with ISTS measurements. Finally, because our conclusions on the increased range of the RKKY interaction are equally valid for a variety of other magnetic objects, such as adatoms, nanoparticles and vacancy-induced magnetic moments, to name but a few, this may open the road to new ways of transporting magnetic information in a material that carries very weak spin-orbit coupling and which thereby dissipates very little spin current.

\begin{acknowledgments}
SRP acknowledges support received from the Irish Research Council for Science, Engineering and Technology under the EMBARK initiative. MSF acknowledges financial support from Science Foundation Ireland under Grant Number SFI 11/RFP.1/MTR/3083. ATC and RBM acknowledge support from INCT Nanocarbono and from CNPq. Computational resources were provided on the Lonsdale cluster maintained by the Trinity Centre for High Performance Computing. This cluster was funded through grants from Science Foundation Ireland. 
\end{acknowledgments}


\begin{thebibliography}{44}
\expandafter\ifx\csname natexlab\endcsname\relax\def\natexlab#1{#1}\fi
\expandafter\ifx\csname bibnamefont\endcsname\relax
  \def\bibnamefont#1{#1}\fi
\expandafter\ifx\csname bibfnamefont\endcsname\relax
  \def\bibfnamefont#1{#1}\fi
\expandafter\ifx\csname citenamefont\endcsname\relax
  \def\citenamefont#1{#1}\fi
\expandafter\ifx\csname url\endcsname\relax
  \def\url#1{\texttt{#1}}\fi
\expandafter\ifx\csname urlprefix\endcsname\relax\def\urlprefix{URL }\fi
\providecommand{\bibinfo}[2]{#2}
\providecommand{\eprint}[2][]{\url{#2}}

\bibitem[{\citenamefont{Castro~Neto et~al.}(2009)\citenamefont{Castro~Neto,
  Guinea, Peres, Novoselov, and Geim}}]{neto:graphrmp}
\bibinfo{author}{\bibfnamefont{A.~H.} \bibnamefont{Castro~Neto}},
  \bibinfo{author}{\bibfnamefont{F.}~\bibnamefont{Guinea}},
  \bibinfo{author}{\bibfnamefont{N.~M.~R.} \bibnamefont{Peres}},
  \bibinfo{author}{\bibfnamefont{K.~S.} \bibnamefont{Novoselov}},
  \bibnamefont{and} \bibinfo{author}{\bibfnamefont{A.~K.} \bibnamefont{Geim}},
  \bibinfo{journal}{Reviews of Modern Physics} \textbf{\bibinfo{volume}{81}},
  \bibinfo{eid}{109} (pages~\bibinfo{numpages}{54}) (\bibinfo{year}{2009}).

\bibitem[{\citenamefont{Geim and Novoselov}(2007)}]{riseofgraphene}
\bibinfo{author}{\bibfnamefont{A.~K.} \bibnamefont{Geim}} \bibnamefont{and}
  \bibinfo{author}{\bibfnamefont{K.~S.} \bibnamefont{Novoselov}},
  \bibinfo{journal}{Nature Materials} \textbf{\bibinfo{volume}{6}},
  \bibinfo{pages}{183} (\bibinfo{year}{2007}).

\bibitem[{\citenamefont{Yazyev}(2010)}]{yazyev:review}
\bibinfo{author}{\bibfnamefont{O.~V.} \bibnamefont{Yazyev}},
  \bibinfo{journal}{Reports on Progress in Physics}
  \textbf{\bibinfo{volume}{73}}, \bibinfo{pages}{056501}
  (\bibinfo{year}{2010}),
  \urlprefix\url{http://stacks.iop.org/0034-4885/73/i=5/a=056501}.

\bibitem[{\citenamefont{Fischer et~al.}(2009)\citenamefont{Fischer, Trauzettel,
  and Loss}}]{fischer_hyperfine_2009}
\bibinfo{author}{\bibfnamefont{J.}~\bibnamefont{Fischer}},
  \bibinfo{author}{\bibfnamefont{B.}~\bibnamefont{Trauzettel}},
  \bibnamefont{and} \bibinfo{author}{\bibfnamefont{D.}~\bibnamefont{Loss}},
  \bibinfo{journal}{Physical Review B} \textbf{\bibinfo{volume}{80}},
  \bibinfo{pages}{155401} (\bibinfo{year}{2009}),
  \urlprefix\url{http://link.aps.org/doi/10.1103/PhysRevB.80.155401}.

\bibitem[{\citenamefont{{Huertas-Hernando}
  et~al.}(2006)\citenamefont{{Huertas-Hernando}, Guinea, and
  Brataas}}]{huertas-hernando_spin-orbit_2006}
\bibinfo{author}{\bibfnamefont{D.}~\bibnamefont{{Huertas-Hernando}}},
  \bibinfo{author}{\bibfnamefont{F.}~\bibnamefont{Guinea}}, \bibnamefont{and}
  \bibinfo{author}{\bibfnamefont{A.}~\bibnamefont{Brataas}},
  \bibinfo{journal}{Physical Review B} \textbf{\bibinfo{volume}{74}},
  \bibinfo{pages}{155426} (\bibinfo{year}{2006}).

\bibitem[{\citenamefont{{Huertas-Hernando}
  et~al.}(2009)\citenamefont{{Huertas-Hernando}, Guinea, and
  Brataas}}]{huertas-hernando_spin-orbit-mediated_2009}
\bibinfo{author}{\bibfnamefont{D.}~\bibnamefont{{Huertas-Hernando}}},
  \bibinfo{author}{\bibfnamefont{F.}~\bibnamefont{Guinea}}, \bibnamefont{and}
  \bibinfo{author}{\bibfnamefont{A.}~\bibnamefont{Brataas}},
  \bibinfo{journal}{Physical Review Letters} \textbf{\bibinfo{volume}{103}},
  \bibinfo{pages}{146801} (\bibinfo{year}{2009}).

\bibitem[{\citenamefont{Kane and Mele}(2005)}]{kane_quantum_2005}
\bibinfo{author}{\bibfnamefont{C.~L.} \bibnamefont{Kane}} \bibnamefont{and}
  \bibinfo{author}{\bibfnamefont{E.~J.} \bibnamefont{Mele}},
  \bibinfo{journal}{Physical Review Letters} \textbf{\bibinfo{volume}{95}},
  \bibinfo{pages}{226801} (\bibinfo{year}{2005}).

\bibitem[{\citenamefont{Min et~al.}(2006)\citenamefont{Min, Hill, Sinitsyn,
  Sahu, Kleinman, and {MacDonald}}}]{min_intrinsic_2006}
\bibinfo{author}{\bibfnamefont{H.}~\bibnamefont{Min}},
  \bibinfo{author}{\bibfnamefont{J.~E.} \bibnamefont{Hill}},
  \bibinfo{author}{\bibfnamefont{N.~A.} \bibnamefont{Sinitsyn}},
  \bibinfo{author}{\bibfnamefont{B.~R.} \bibnamefont{Sahu}},
  \bibinfo{author}{\bibfnamefont{L.}~\bibnamefont{Kleinman}}, \bibnamefont{and}
  \bibinfo{author}{\bibfnamefont{A.~H.} \bibnamefont{{MacDonald}}},
  \bibinfo{journal}{Physical Review B} \textbf{\bibinfo{volume}{74}},
  \bibinfo{pages}{165310} (\bibinfo{year}{2006}).

\bibitem[{\citenamefont{Trauzettel et~al.}(2007)\citenamefont{Trauzettel,
  Bulaev, Loss, and Burkard}}]{trauzettel_spin_2007}
\bibinfo{author}{\bibfnamefont{B.}~\bibnamefont{Trauzettel}},
  \bibinfo{author}{\bibfnamefont{D.~V.} \bibnamefont{Bulaev}},
  \bibinfo{author}{\bibfnamefont{D.}~\bibnamefont{Loss}}, \bibnamefont{and}
  \bibinfo{author}{\bibfnamefont{G.}~\bibnamefont{Burkard}},
  \bibinfo{journal}{Nature Physics} \textbf{\bibinfo{volume}{3}},
  \bibinfo{pages}{192} (\bibinfo{year}{2007}), ISSN \bibinfo{issn}{1745-2473},
  \urlprefix\url{http://dx.doi.org/10.1038/nphys544}.

\bibitem[{\citenamefont{Yazyev}(2008)}]{yazyev_hyperfine_2008}
\bibinfo{author}{\bibfnamefont{O.~V.} \bibnamefont{Yazyev}},
  \bibinfo{journal}{Nano Letters} \textbf{\bibinfo{volume}{8}},
  \bibinfo{pages}{1011} (\bibinfo{year}{2008}), ISSN \bibinfo{issn}{1530-6984},
  \urlprefix\url{http://dx.doi.org/10.1021/nl072667q}.

\bibitem[{\citenamefont{Kasuya}(1956)}]{RKKY:K}
\bibinfo{author}{\bibfnamefont{T.}~\bibnamefont{Kasuya}},
  \bibinfo{journal}{Progress of Theoretical Physics}
  \textbf{\bibinfo{volume}{16}}, \bibinfo{pages}{45} (\bibinfo{year}{1956}).

\bibitem[{\citenamefont{Ruderman and Kittel}(1954)}]{RKKY:RK}
\bibinfo{author}{\bibfnamefont{M.~A.} \bibnamefont{Ruderman}} \bibnamefont{and}
  \bibinfo{author}{\bibfnamefont{C.}~\bibnamefont{Kittel}},
  \bibinfo{journal}{Physical Review} \textbf{\bibinfo{volume}{96}},
  \bibinfo{pages}{99} (\bibinfo{year}{1954}).

\bibitem[{\citenamefont{Yosida}(1957)}]{RKKY:Y}
\bibinfo{author}{\bibfnamefont{K.}~\bibnamefont{Yosida}},
  \bibinfo{journal}{Physical Review} \textbf{\bibinfo{volume}{106}},
  \bibinfo{pages}{893} (\bibinfo{year}{1957}).

\bibitem[{\citenamefont{{Black-Schaffer}}(2010)}]{black-schaffer_importance_20%
10}
\bibinfo{author}{\bibfnamefont{A.~M.} \bibnamefont{{Black-Schaffer}}},
  \bibinfo{journal}{Physical Review B} \textbf{\bibinfo{volume}{82}},
  \bibinfo{pages}{073409} (\bibinfo{year}{2010}).

\bibitem[{\citenamefont{Black-Schaffer}(2010)}]{black:graphenerkky}
\bibinfo{author}{\bibfnamefont{A.~M.} \bibnamefont{Black-Schaffer}},
  \bibinfo{journal}{Physical Review B} \textbf{\bibinfo{volume}{81}},
  \bibinfo{pages}{205416} (\bibinfo{year}{2010}).

\bibitem[{\citenamefont{Brey et~al.}(2007)\citenamefont{Brey, Fertig, and
  Sarma}}]{brey:graphenerkky}
\bibinfo{author}{\bibfnamefont{L.}~\bibnamefont{Brey}},
  \bibinfo{author}{\bibfnamefont{H.~A.} \bibnamefont{Fertig}},
  \bibnamefont{and} \bibinfo{author}{\bibfnamefont{S.~D.} \bibnamefont{Sarma}},
  \bibinfo{journal}{Physical Review Letters} \textbf{\bibinfo{volume}{99}},
  \bibinfo{eid}{116802} (pages~\bibinfo{numpages}{4}) (\bibinfo{year}{2007}).

\bibitem[{\citenamefont{Bunder and Lin}(2009)}]{bunder:rkkygraphene}
\bibinfo{author}{\bibfnamefont{J.~E.} \bibnamefont{Bunder}} \bibnamefont{and}
  \bibinfo{author}{\bibfnamefont{H.}~\bibnamefont{Lin}},
  \bibinfo{journal}{Physical Review B} \textbf{\bibinfo{volume}{80}},
  \bibinfo{pages}{153414} (\bibinfo{year}{2009}),
  \urlprefix\url{http://link.aps.org/doi/10.1103/PhysRevB.80.153414}.

\bibitem[{\citenamefont{Dugaev et~al.}(2006)\citenamefont{Dugaev, Litvinov, and
  Barnas}}]{dugaev:rkkygraphene}
\bibinfo{author}{\bibfnamefont{V.~K.} \bibnamefont{Dugaev}},
  \bibinfo{author}{\bibfnamefont{V.~I.} \bibnamefont{Litvinov}},
  \bibnamefont{and} \bibinfo{author}{\bibfnamefont{J.}~\bibnamefont{Barnas}},
  \bibinfo{journal}{Physical Review B} \textbf{\bibinfo{volume}{74}},
  \bibinfo{pages}{224438} (\bibinfo{year}{2006}),
  \urlprefix\url{http://link.aps.org/doi/10.1103/PhysRevB.74.224438}.

\bibitem[{\citenamefont{Hwang and Das~Sarma}(2008)}]{hwang:rkkygraphene}
\bibinfo{author}{\bibfnamefont{E.~H.} \bibnamefont{Hwang}} \bibnamefont{and}
  \bibinfo{author}{\bibfnamefont{S.}~\bibnamefont{Das~Sarma}},
  \bibinfo{journal}{Physical Review Letters} \textbf{\bibinfo{volume}{101}},
  \bibinfo{pages}{156802} (\bibinfo{year}{2008}),
  \urlprefix\url{http://link.aps.org/doi/10.1103/PhysRevLett.101.156802}.

\bibitem[{\citenamefont{Kogan}(2011)}]{kogan:rkkygraphene}
\bibinfo{author}{\bibfnamefont{E.}~\bibnamefont{Kogan}},
  \bibinfo{journal}{Physical Review B} \textbf{\bibinfo{volume}{84}},
  \bibinfo{pages}{115119} (\bibinfo{year}{2011}),
  \urlprefix\url{http://link.aps.org/doi/10.1103/PhysRevB.84.115119}.

\bibitem[{\citenamefont{Lee et~al.}(2011)\citenamefont{Lee, Kim, Mucciolo,
  Bouzerar, and Kettemann}}]{disorderedRKKY}
\bibinfo{author}{\bibfnamefont{H.}~\bibnamefont{Lee}},
  \bibinfo{author}{\bibfnamefont{J.}~\bibnamefont{Kim}},
  \bibinfo{author}{\bibfnamefont{E.~R.} \bibnamefont{Mucciolo}},
  \bibinfo{author}{\bibfnamefont{G.}~\bibnamefont{Bouzerar}}, \bibnamefont{and}
  \bibinfo{author}{\bibfnamefont{S.}~\bibnamefont{Kettemann}},
  \bibinfo{journal}{{arXiv:1110.6272}}  (\bibinfo{year}{2011}),
  \urlprefix\url{http://arxiv.org/abs/1110.6272}.

\bibitem[{\citenamefont{Power and Ferreira}(2011)}]{me:grapheneGF}
\bibinfo{author}{\bibfnamefont{S.~R.} \bibnamefont{Power}} \bibnamefont{and}
  \bibinfo{author}{\bibfnamefont{M.~S.} \bibnamefont{Ferreira}},
  \bibinfo{journal}{Physical Review B} \textbf{\bibinfo{volume}{83}},
  \bibinfo{pages}{155432} (\bibinfo{year}{2011}),
  \urlprefix\url{http://link.aps.org/doi/10.1103/PhysRevB.83.155432}.

\bibitem[{\citenamefont{Saremi}(2007)}]{saremi:graphenerkky}
\bibinfo{author}{\bibfnamefont{S.}~\bibnamefont{Saremi}},
  \bibinfo{journal}{Physical Review B} \textbf{\bibinfo{volume}{76}},
  \bibinfo{eid}{184430} (pages~\bibinfo{numpages}{6}) (\bibinfo{year}{2007}).

\bibitem[{\citenamefont{Sherafati and
  Satpathy}(2011{\natexlab{a}})}]{sherafati:rkkygraphene2}
\bibinfo{author}{\bibfnamefont{M.}~\bibnamefont{Sherafati}} \bibnamefont{and}
  \bibinfo{author}{\bibfnamefont{S.}~\bibnamefont{Satpathy}},
  \bibinfo{journal}{Physical Review B} \textbf{\bibinfo{volume}{84}},
  \bibinfo{pages}{125416} (\bibinfo{year}{2011}{\natexlab{a}}),
  \urlprefix\url{http://link.aps.org/doi/10.1103/PhysRevB.84.125416}.

\bibitem[{\citenamefont{Sherafati and
  Satpathy}(2011{\natexlab{b}})}]{sherafati:graphenerkky}
\bibinfo{author}{\bibfnamefont{M.}~\bibnamefont{Sherafati}} \bibnamefont{and}
  \bibinfo{author}{\bibfnamefont{S.}~\bibnamefont{Satpathy}},
  \bibinfo{journal}{Physical Review B} \textbf{\bibinfo{volume}{83}},
  \bibinfo{pages}{165425} (\bibinfo{year}{2011}{\natexlab{b}}).

\bibitem[{\citenamefont{Uchoa et~al.}(2011)\citenamefont{Uchoa, Rappoport, and
  Castro~Neto}}]{uchoa:rkkygraphene}
\bibinfo{author}{\bibfnamefont{B.}~\bibnamefont{Uchoa}},
  \bibinfo{author}{\bibfnamefont{T.~G.} \bibnamefont{Rappoport}},
  \bibnamefont{and} \bibinfo{author}{\bibfnamefont{A.~H.}
  \bibnamefont{Castro~Neto}}, \bibinfo{journal}{Physical Review Letters}
  \textbf{\bibinfo{volume}{106}}, \bibinfo{pages}{016801}
  (\bibinfo{year}{2011}),
  \urlprefix\url{http://link.aps.org/doi/10.1103/PhysRevLett.106.016801}.

\bibitem[{\citenamefont{Vozmediano et~al.}(2005)\citenamefont{Vozmediano,
  L\'opez-Sancho, Stauber, and Guinea}}]{Vozmediano:2005}
\bibinfo{author}{\bibfnamefont{M.~A.~H.} \bibnamefont{Vozmediano}},
  \bibinfo{author}{\bibfnamefont{M.~P.} \bibnamefont{L\'opez-Sancho}},
  \bibinfo{author}{\bibfnamefont{T.}~\bibnamefont{Stauber}}, \bibnamefont{and}
  \bibinfo{author}{\bibfnamefont{F.}~\bibnamefont{Guinea}},
  \bibinfo{journal}{Physical Review B} \textbf{\bibinfo{volume}{72}},
  \bibinfo{pages}{155121} (\bibinfo{year}{2005}).

\bibitem[{\citenamefont{Krstić et~al.}(2010)\citenamefont{Krstić, Ewels,
  W\aa{}gberg, Ferreira, Janssens, Stéphan, and Glerup}}]{vojislav:acs}
\bibinfo{author}{\bibfnamefont{V.}~\bibnamefont{Krstić}},
  \bibinfo{author}{\bibfnamefont{C.~P.} \bibnamefont{Ewels}},
  \bibinfo{author}{\bibfnamefont{T.}~\bibnamefont{W\aa{}gberg}},
  \bibinfo{author}{\bibfnamefont{M.~S.} \bibnamefont{Ferreira}},
  \bibinfo{author}{\bibfnamefont{A.~M.} \bibnamefont{Janssens}},
  \bibinfo{author}{\bibfnamefont{O.}~\bibnamefont{Stéphan}},
  \bibnamefont{and} \bibinfo{author}{\bibfnamefont{M.}~\bibnamefont{Glerup}},
  \bibinfo{journal}{ACS Nano} \textbf{\bibinfo{volume}{4}},
  \bibinfo{pages}{5081} (\bibinfo{year}{2010}),
  \eprint{http://pubs.acs.org/doi/pdf/10.1021/nn1009038},
  \urlprefix\url{http://pubs.acs.org/doi/abs/10.1021/nn1009038}.

\bibitem[{\citenamefont{Costa et~al.}(2005)\citenamefont{Costa, Kirwan, and
  Ferreira}}]{AntonioDavidIEC}
\bibinfo{author}{\bibfnamefont{A.~T.} \bibnamefont{Costa}},
  \bibinfo{author}{\bibfnamefont{D.~F.} \bibnamefont{Kirwan}},
  \bibnamefont{and} \bibinfo{author}{\bibfnamefont{M.~S.}
  \bibnamefont{Ferreira}}, \bibinfo{journal}{Physical Review B}
  \textbf{\bibinfo{volume}{72}}, \bibinfo{pages}{085402}
  (\bibinfo{year}{2005}).

\bibitem[{\citenamefont{Kirwan et~al.}(2009)\citenamefont{Kirwan, de~Menezes,
  Rocha, Costa, Muniz, Fagan, and Ferreira}}]{DavidSpinValve}
\bibinfo{author}{\bibfnamefont{D.~F.} \bibnamefont{Kirwan}},
  \bibinfo{author}{\bibfnamefont{V.~M.} \bibnamefont{de~Menezes}},
  \bibinfo{author}{\bibfnamefont{C.~G.} \bibnamefont{Rocha}},
  \bibinfo{author}{\bibfnamefont{A.~T.} \bibnamefont{Costa}},
  \bibinfo{author}{\bibfnamefont{R.~B.} \bibnamefont{Muniz}},
  \bibinfo{author}{\bibfnamefont{S.~B.} \bibnamefont{Fagan}}, \bibnamefont{and}
  \bibinfo{author}{\bibfnamefont{M.~S.} \bibnamefont{Ferreira}},
  \bibinfo{journal}{Carbon} \textbf{\bibinfo{volume}{47}},
  \bibinfo{pages}{2533} (\bibinfo{year}{2009}).

\bibitem[{\citenamefont{Kirwan et~al.}(2008)\citenamefont{Kirwan, Rocha, Costa,
  and Ferreira}}]{David:IEC}
\bibinfo{author}{\bibfnamefont{D.~F.} \bibnamefont{Kirwan}},
  \bibinfo{author}{\bibfnamefont{C.~G.} \bibnamefont{Rocha}},
  \bibinfo{author}{\bibfnamefont{A.~T.} \bibnamefont{Costa}}, \bibnamefont{and}
  \bibinfo{author}{\bibfnamefont{M.~S.} \bibnamefont{Ferreira}},
  \bibinfo{journal}{Physical Review B} \textbf{\bibinfo{volume}{77}},
  \bibinfo{eid}{085432} (pages~\bibinfo{numpages}{6}) (\bibinfo{year}{2008}).

\bibitem[{obs()}]{obs}
\bibinfo{note}{Undoped graphene represents graphene with its Fermi level at
  half filling.}

\bibitem[{\citenamefont{\v{S}im\'anek and
  Heinrich}(2003)}]{Simanek_gilbert_2003}
\bibinfo{author}{\bibfnamefont{E.}~\bibnamefont{\v{S}im\'anek}}
  \bibnamefont{and} \bibinfo{author}{\bibfnamefont{B.}~\bibnamefont{Heinrich}},
  \bibinfo{journal}{Physical Review B} \textbf{\bibinfo{volume}{67}},
  \bibinfo{pages}{144418} (\bibinfo{year}{2003}),
  \urlprefix\url{http://link.aps.org/doi/10.1103/PhysRevB.67.144418}.

\bibitem[{\citenamefont{Heinrich et~al.}(2003)\citenamefont{Heinrich,
  Tserkovnyak, Woltersdorf, Brataas, Urban, and Bauer}}]{heinrich_dynamic_2003}
\bibinfo{author}{\bibfnamefont{B.}~\bibnamefont{Heinrich}},
  \bibinfo{author}{\bibfnamefont{Y.}~\bibnamefont{Tserkovnyak}},
  \bibinfo{author}{\bibfnamefont{G.}~\bibnamefont{Woltersdorf}},
  \bibinfo{author}{\bibfnamefont{A.}~\bibnamefont{Brataas}},
  \bibinfo{author}{\bibfnamefont{R.}~\bibnamefont{Urban}}, \bibnamefont{and}
  \bibinfo{author}{\bibfnamefont{G.~E.~W.} \bibnamefont{Bauer}},
  \bibinfo{journal}{Physical Review Letters} \textbf{\bibinfo{volume}{90}},
  \bibinfo{pages}{187601} (\bibinfo{year}{2003}),
  \urlprefix\url{http://link.aps.org/doi/10.1103/PhysRevLett.90.187601}.

\bibitem[{\citenamefont{Costache et~al.}(2008)\citenamefont{Costache, Watts,
  van~der Wal, and van Wees}}]{costache_electrical_2008}
\bibinfo{author}{\bibfnamefont{M.~V.} \bibnamefont{Costache}},
  \bibinfo{author}{\bibfnamefont{S.~M.} \bibnamefont{Watts}},
  \bibinfo{author}{\bibfnamefont{C.~H.} \bibnamefont{van~der Wal}},
  \bibnamefont{and} \bibinfo{author}{\bibfnamefont{B.~J.} \bibnamefont{van
  Wees}}, \bibinfo{journal}{Physical Review B} \textbf{\bibinfo{volume}{78}},
  \bibinfo{pages}{064423} (\bibinfo{year}{2008}),
  \urlprefix\url{http://link.aps.org/doi/10.1103/PhysRevB.78.064423}.

\bibitem[{\citenamefont{Tombros et~al.}(2007)\citenamefont{Tombros, Jozsa,
  Popinciuc, Jonkman, and van Wees}}]{tombros_electronic_2007}
\bibinfo{author}{\bibfnamefont{N.}~\bibnamefont{Tombros}},
  \bibinfo{author}{\bibfnamefont{C.}~\bibnamefont{Jozsa}},
  \bibinfo{author}{\bibfnamefont{M.}~\bibnamefont{Popinciuc}},
  \bibinfo{author}{\bibfnamefont{H.~T.} \bibnamefont{Jonkman}},
  \bibnamefont{and} \bibinfo{author}{\bibfnamefont{B.~J.} \bibnamefont{van
  Wees}}, \bibinfo{journal}{Nature} \textbf{\bibinfo{volume}{448}},
  \bibinfo{pages}{571} (\bibinfo{year}{2007}), ISSN \bibinfo{issn}{0028-0836},
  \urlprefix\url{http://dx.doi.org/10.1038/nature06037}.

\bibitem[{\citenamefont{Tserkovnyak et~al.}(2005)\citenamefont{Tserkovnyak,
  Brataas, Bauer, and Halperin}}]{tserkovnyak_nonlocal_2005}
\bibinfo{author}{\bibfnamefont{Y.}~\bibnamefont{Tserkovnyak}},
  \bibinfo{author}{\bibfnamefont{A.}~\bibnamefont{Brataas}},
  \bibinfo{author}{\bibfnamefont{G.~E.~W.} \bibnamefont{Bauer}},
  \bibnamefont{and} \bibinfo{author}{\bibfnamefont{B.~I.}
  \bibnamefont{Halperin}}, \bibinfo{journal}{Reviews of Modern Physics}
  \textbf{\bibinfo{volume}{77}}, \bibinfo{pages}{1375} (\bibinfo{year}{2005}),
  \urlprefix\url{http://link.aps.org/doi/10.1103/RevModPhys.77.1375}.

\bibitem[{\citenamefont{Anderson}(1961)}]{anderson_localized_1961}
\bibinfo{author}{\bibfnamefont{P.~W.} \bibnamefont{Anderson}},
  \bibinfo{journal}{Physical Review} \textbf{\bibinfo{volume}{124}},
  \bibinfo{pages}{41} (\bibinfo{year}{1961}),
  \urlprefix\url{http://link.aps.org/doi/10.1103/PhysRev.124.41}.

\bibitem[{\citenamefont{Muniz and Mills}(2002)}]{muniz_theory_2002}
\bibinfo{author}{\bibfnamefont{R.~B.} \bibnamefont{Muniz}} \bibnamefont{and}
  \bibinfo{author}{\bibfnamefont{D.~L.} \bibnamefont{Mills}},
  \bibinfo{journal}{Physical Review B} \textbf{\bibinfo{volume}{66}},
  \bibinfo{pages}{174417} (\bibinfo{year}{2002}),
  \urlprefix\url{http://link.aps.org/doi/10.1103/PhysRevB.66.174417}.

\bibitem[{\citenamefont{Sielemann et~al.}(2008)\citenamefont{Sielemann,
  Kobayashi, Yoshida, Gunnlaugsson, and Weyer}}]{sielemann_magnetism_2008}
\bibinfo{author}{\bibfnamefont{R.}~\bibnamefont{Sielemann}},
  \bibinfo{author}{\bibfnamefont{Y.}~\bibnamefont{Kobayashi}},
  \bibinfo{author}{\bibfnamefont{Y.}~\bibnamefont{Yoshida}},
  \bibinfo{author}{\bibfnamefont{H.~P.} \bibnamefont{Gunnlaugsson}},
  \bibnamefont{and} \bibinfo{author}{\bibfnamefont{G.}~\bibnamefont{Weyer}},
  \bibinfo{journal}{Physical Review Letters} \textbf{\bibinfo{volume}{101}},
  \bibinfo{pages}{137206} (\bibinfo{year}{2008}),
  \urlprefix\url{http://link.aps.org/doi/10.1103/PhysRevLett.101.137206}.

\bibitem[{\citenamefont{Heinrich et~al.}(2004)\citenamefont{Heinrich, Gupta,
  Lutz, and Eigler}}]{heinrich_single-atom_2004}
\bibinfo{author}{\bibfnamefont{A.~J.} \bibnamefont{Heinrich}},
  \bibinfo{author}{\bibfnamefont{J.~A.} \bibnamefont{Gupta}},
  \bibinfo{author}{\bibfnamefont{C.~P.} \bibnamefont{Lutz}}, \bibnamefont{and}
  \bibinfo{author}{\bibfnamefont{D.~M.} \bibnamefont{Eigler}},
  \bibinfo{journal}{Science} \textbf{\bibinfo{volume}{306}},
  \bibinfo{pages}{466} (\bibinfo{year}{2004}),
  \urlprefix\url{http://www.sciencemag.org/content/306/5695/466.abstract}.

\bibitem[{\citenamefont{Khajetoorians et~al.}(2010)\citenamefont{Khajetoorians,
  Chilian, Wiebe, Schuwalow, Lechermann, and
  Wiesendanger}}]{khajetoorians_detecting_2010}
\bibinfo{author}{\bibfnamefont{A.~A.} \bibnamefont{Khajetoorians}},
  \bibinfo{author}{\bibfnamefont{B.}~\bibnamefont{Chilian}},
  \bibinfo{author}{\bibfnamefont{J.}~\bibnamefont{Wiebe}},
  \bibinfo{author}{\bibfnamefont{S.}~\bibnamefont{Schuwalow}},
  \bibinfo{author}{\bibfnamefont{F.}~\bibnamefont{Lechermann}},
  \bibnamefont{and}
  \bibinfo{author}{\bibfnamefont{R.}~\bibnamefont{Wiesendanger}},
  \bibinfo{journal}{Nature} \textbf{\bibinfo{volume}{467}},
  \bibinfo{pages}{1084} (\bibinfo{year}{2010}), ISSN \bibinfo{issn}{0028-0836},
  \urlprefix\url{http://dx.doi.org/10.1038/nature09519}.

\bibitem[{\citenamefont{Khajetoorians et~al.}(2011)\citenamefont{Khajetoorians,
  Lounis, Chilian, Costa, Zhou, Mills, Wiebe, and
  Wiesendanger}}]{khajetoorians_itinerant_2011}
\bibinfo{author}{\bibfnamefont{A.~A.} \bibnamefont{Khajetoorians}},
  \bibinfo{author}{\bibfnamefont{S.}~\bibnamefont{Lounis}},
  \bibinfo{author}{\bibfnamefont{B.}~\bibnamefont{Chilian}},
  \bibinfo{author}{\bibfnamefont{A.~T.} \bibnamefont{Costa}},
  \bibinfo{author}{\bibfnamefont{L.}~\bibnamefont{Zhou}},
  \bibinfo{author}{\bibfnamefont{D.~L.} \bibnamefont{Mills}},
  \bibinfo{author}{\bibfnamefont{J.}~\bibnamefont{Wiebe}}, \bibnamefont{and}
  \bibinfo{author}{\bibfnamefont{R.}~\bibnamefont{Wiesendanger}},
  \bibinfo{journal}{Physical Review Letters} \textbf{\bibinfo{volume}{106}},
  \bibinfo{pages}{037205} (\bibinfo{year}{2011}).

\bibitem[{\citenamefont{Loth et~al.}(2010)\citenamefont{Loth, von Bergmann,
  Ternes, Otte, Lutz, and Heinrich}}]{loth_controlling_2010}
\bibinfo{author}{\bibfnamefont{S.}~\bibnamefont{Loth}},
  \bibinfo{author}{\bibfnamefont{K.}~\bibnamefont{von Bergmann}},
  \bibinfo{author}{\bibfnamefont{M.}~\bibnamefont{Ternes}},
  \bibinfo{author}{\bibfnamefont{A.~F.} \bibnamefont{Otte}},
  \bibinfo{author}{\bibfnamefont{C.~P.} \bibnamefont{Lutz}}, \bibnamefont{and}
  \bibinfo{author}{\bibfnamefont{A.~J.} \bibnamefont{Heinrich}},
  \bibinfo{journal}{Nature Physics} \textbf{\bibinfo{volume}{6}},
  \bibinfo{pages}{340} (\bibinfo{year}{2010}), ISSN \bibinfo{issn}{1745-2473},
  \urlprefix\url{http://dx.doi.org/10.1038/nphys1616}.

\end{thebibliography}
\end{document}